# Weak localization of electromagnetic waves and radar polarimetry of Saturn's rings

Michael I. Mishchenko[1*] and Janna M. Dlugach[2]

[1] *NASA Goddard Institute for Space Studies, 2880 Broadway, New York, NY 10025, USA*
[2] *Main Astronomical Observatory of the National Academy of Sciences of Ukraine, 27 Zabolotny Str., 03680, Kyiv, Ukraine*

**ABSTRACT**

We use a state-of-the-art physics-based model of electromagnetic scattering to analyze average circular polarization ratios measured for the A and B rings of Saturn at a wavelength of 12.6 cm. This model is directly based on the Maxwell equations and accounts for the effects of polarization, multiple scattering, weak localization of electromagnetic waves, and ring particle nonsphericity. Our analysis is based on the assumption that the observed polarization ratios are accurate, mutually consistent, and show a quasi-linear dependence on the opening angle. Also, we assume that the ring system is not strongly stratified in the vertical direction. Our numerical simulations rule out the model of spherical ring particles, favor the model of ring bodies in the form of nearly spherical particles with small-scale surface roughness, and rule out nonspherical particles with aspect ratios significantly exceeding 1.2. They also favor particles with effective radii in the range 4–10 cm and definitely rule out effective radii significantly smaller than 4 cm. Furthermore, they seem to rule out effective radii significantly greater than 10 cm. The retrieved ring optical thickness values are in the range 2–3 or even larger. If the rings do have a wake-like horizontal structure, as has been recently suggested, then these optical thickness values should be attributed to an average wake rather than to the optical thickness averaged over the entire horizontal extent of the rings.

[*]E-mail: mmishchenko@giss.nasa.gov

## 1 INTRODUCTION

Radar observations have played a very important role in the evolving study of the physical nature of the A and B rings of Saturn (Goldstein et al. 1977; Ostro 1993; Nicholson et al. 2005). Polarization radar measurements have always been expected to be especially indicative of the physical properties of the ring particles. Indeed, while absolute measurements of the radar cross sections may suffer from several sources of errors and uncertainties, most of systematic uncertainties cancel out in the computation of the circular polarization ratio $\mu_C$, thereby resulting in a more reliably determined quantity (Ostro 1993; Nicholson et al. 2005). Also, by virtue of being the result of dividing one intensity by another, $\mu_\text{C}$ can be expected to be significantly less affected by the horizontal and vertical inhomogeneity of the rings, e.g., by the potential wake structure (Daisaka et al. 2001). Furthermore, while the existing measurements of the radar cross sections published by different authors are somewhat inconsistent (cf. Nicholson et al. 2005), the measurements of circular polarization at a wavelength of 12.6 cm by Ostro et al. (1980) and Nicholson et al. (2005) appear to be quite consistent and show a systematic and nearly linear increase in $\mu_\text{C}$ with increasing ring opening angle $B$ (the angle between the line of sight and the ring plane).

This quasi-linear angular trend in $\mu_\text{C}$, if real, is somewhat puzzling and has so far defied a physically-based quantitative explanation. The main reason for this is the substantial theoretical and numerical complexity of the inverse remote-sensing problem. Indeed, to perform a quantitative analysis of radar depolarization measurements, one must fully take into account the effects of polarization, multiple scattering, weak localization of electromagnetic waves, and ring particle nonsphericity. To the best of our knowledge, this has not been done before and is, therefore, the main objective of this paper.

The scope of this study is intentionally limited. Specifically, we assume that the quasi-linear angular trend in the circular polarization ratio at 12.6 cm is real, use a vertically homogeneous many-particle-thick model of the A and B rings, and analyze the measurements using a state-of-the-art physics-based theoretical approach. We show that the radar polarization measurements alone impose rather strong limitations on the potential range of ring particle models and on the optical thickness of the rings. A combined analysis of all radar and occultation observations of Saturn's rings may require a refined model of the ring structure and can be expected to further narrow the possible range of solutions. However, we believe that it may also require a better prior understanding of the degree of consistency of the various data sets and, possibly, improved analysis techniques.

The following section provides a brief description of the electromagnetic scattering model used in this study. The results of numerical computations are presented in Section 3 and analyzed in Section 4.

## 2  THEORY AND NUMERICAL TECHNIQUES

The average $\mu_\text{C}$ values measured for the A and B rings and the corresponding error bars are shown in Fig. 1. The strong dependence of $\mu_\text{C}$ on the ring opening angle is obviously indicative of a significant effect of multiple scattering and, thus, rules out the monolayer model of the rings. It also rules out strongly depolarizing individual ring particles since they would cause large polarization ratios at $B \approx 0°$. Indeed, in the case of grazing incidence and reflection, the backscattered signal is caused almost entirely by the first-order scattering even in a many-particle-thick ring system (Hovenier & Stam 2007). Therefore, the extrapolation of the measurement results to $B = 0°$ (Fig. 1) is obviously indicative of weakly depolarizing ring particles. We thus need to use a model of the rings which is at least several particles thick and thereby supports substantial multiple scattering. Furthermore, the particles must be expressly nonspherical because it is unrealistic to expect that somehow all ring particles have acquired a perfect spherical shape.

At a very large distance from the antenna, the transmitted electromagnetic wave becomes locally plane. The total scattered field at the observation point can always be represented as a superposition of partial contributions corresponding to every possible sequence of ring particles (Mishchenko 2008):

$$\mathbf{E}^{\mathrm{sca}}(\mathbf{r}) = \sum_{i=1}^{\infty} \widetilde{\mathbf{E}}_i(\mathbf{r}), \qquad (1)$$

where $\mathbf{r}$ is the position vector of the observation point and the index $i$ numbers the particle sequences. A sequence can consist of just one particle or of two or more particles in a specific order. At a distant observation point, the partial field due to any particle sequence contributing to the right-hand side of Eq. (1) becomes an outgoing spherical wavelet centered at the last particle of the sequence. This occurs irrespective of whether the particles are densely packed or sparsely distributed. The Stokes parameters of the scattered light can be directly expressed in terms of the elements of the so-called scattering coherency dyad $\vec{\rho}^{\mathrm{sca}} = \mathbf{E}^{\mathrm{sca}} \otimes (\mathbf{E}^{\mathrm{sca}})^*$, where the asterisk denotes complex conjugation and $\otimes$ is the dyadic product sign. The dyadic product of the right-hand side of Eq. (1) with its complex-conjugate counterpart is the sum of an infinite number of terms, each describing the result of interference of two spherical wavelets centered at the end particles of a pair of particle sequences:

$$\vec{\rho}^{\mathrm{sca}} = \sum_{i=1}^{\infty} \sum_{j=1}^{\infty} \widetilde{\mathbf{E}}_i(\mathbf{r}) \otimes [\widetilde{\mathbf{E}}_j(\mathbf{r})]^*. \qquad (2)$$

Thus the scattering signal detected by the receiving antenna is the cumulative result of interference of pairs of wavelets generated by various sequences of particles. Three typical cases of wavelet pairs are shown in Fig. 2, in which $\hat{\mathbf{n}}^{\mathrm{inc}}$ is the unit vector in the incidence direction and $\hat{\mathbf{n}}^{\mathrm{sca}}$ is the unit vector in the scattering direction. The result of interference of the two wavelets shown in Fig. 2a depends on the phase difference between the wavelets, which changes rapidly with changing particle positions during the measurement. Such pairs of wavelets create a rapidly varying speckle pattern which is completely averaged out by the accumulation of the scattering signal over a period of time (Mishchenko 2008).

In contrast, the two wavelets shown in Fig. 2b are identical since both of them are created by the same group of particles taken in the same order. Therefore, the phase difference between the wavelets is identically equal to zero, thereby resulting in consistently constructive self-interference. The wavelet pairs of this type create the diffuse radiation background described by the vector radiative transfer equation (VRTE) provided that the number of particles in the scattering medium is very large and their packing density is sufficiently low (Mishchenko et al. 2006; Mishchenko 2008).

The type of wavelet pairs shown in Fig. 2c is special in that both wavelets are created by the same group of particles but taken in the opposite order. The average result of interference of such conjugate wavelets is equal to zero in all scattering directions except very close to the exact backscattering direction given by $\hat{\mathbf{n}}^{\text{sca}} = -\hat{\mathbf{n}}^{\text{inc}}$. In this latter case the phase difference between the wavelets vanishes, thereby causing a pronounced backscattering intensity peak. This effect is called weak localization of electromagnetic waves or coherent backscattering (Barabanenkov et al. 1991; Lenke & Maret 2000; Mishchenko et al. 2006). Although the computation of the angular profile of weak localization for a many-particle group is an exceedingly complex problem (Muinonen 2004; Videen et al. 2004; Tishkovets and Jockers 2006; Litvinov et al. 2007; Tishkovets 2007), it has been shown that in the exact backscattering direction all characteristics of weak localization can be rigorously expressed in terms of the solution of the VRTE (Mishchenko 1991). This result is very important since it corresponds precisely to the case of monostatic radar observations.

Thus the theoretical modeling of polarized radar returns involves the following consecutive steps:

1. the computation of the relevant ensemble-averaged single-scattering properties of the ring particles;
2. the computation of the diffuse Stokes reflection matrix for a homogeneous plane-parallel model of Saturn's rings through the explicit numerical solution of the VRTE;
3. the computation of the requisite characteristics of weak localization in the exact backscattering direction from the diffuse Stokes reflection matrix; and, finally,
4. the computation of the circular polarization ratio $\mu_{\text{C}}$.

The entire procedure is described in exquisite detail in Mishchenko (1996) and Mishchenko et al. (2002, 2006). Below we highlight only those modeling aspects that are specific to this study.

The single-scattering characteristics of polydisperse nonspherical particles are computed using the numerically exact *T*-matrix method (Waterman 1971; Mishchenko et al. 2002). Figure 3 illustrates the particle models used in this study. The shape of a rotationally symmetric so-called Chebyshev particle with respect to the particle reference frame is given by $R(\theta) = r_0(1 + \xi \cos n\theta)$, where $\theta$ is the polar angle, $r_0$ is the radius of the unperturbed sphere, $n$ is the waviness parameter, and $\xi$ is the deformation parameter (Wiscombe & Mugnai 1986). The latter can be either negative or positive. Chebyshev particles with $|\xi| < 0.15$ are moderately aspherical and exhibit surface roughness in the form of waves running completely around the particle. The number of waves is proportional to *n*. In all our computations the value of the waviness parameter was fixed at 6. Although this choice might appear to be somewhat arbitrary, it serves our objectives well in that it is both small enough to allow for numerically converging *T*-matrix computations for sufficiently large particles and large enough to yield a distinctly undulating surface of an otherwise compact particle. To analyze the effects of a major deviation of the particle shape from that of a perfect sphere, we also use the model of oblate and prolate spheroids. The shape of a spheroid is specified in terms of the semi-axis ratio $a/b$, where $b$ is the spheroid semi-axis along the axis of rotation and $a$ is the semi-axis in the perpendicular direction.

In all computations we assume that the aspect ratio (or the average aspect ratio) of spheroids $\varepsilon$ is 1.5, where $\varepsilon$ is defined as the ratio of the maximal to the minimal dimension of a particle.

The nonspherical ring particles are assumed to be randomly oriented. The size of each particle is specified in terms of the radius $r$ of the sphere having the same surface area. The probability distribution of the equivalent-sphere radii is assumed to follow the simple power law:

$$n(r) = \begin{cases} \text{constant} \times r^{-3}, & r_1 \leq r \leq r_2, \\ 0, & \text{otherwise}, \end{cases} \qquad (3)$$

the constant being determined from the standard normalization condition

$$\int_{r_1}^{r_2} dr\, n(r) = 1. \qquad (4)$$

Instead of specifying the minimal and maximal radii $r_1$ and $r_2$, the size distribution is characterized in terms of the effective radius and effective variance defined by

$$r_{\text{eff}} = \frac{1}{\langle G \rangle} \int_{r_1}^{r_2} dr\, n(r) r \pi r^2 = \frac{r_2 - r_1}{\ln(r_2/r_1)} \qquad (5)$$

and

$$v_{\text{eff}} = \frac{1}{\langle G \rangle r_{\text{eff}}^2} \int_{r_1}^{r_2} dr\, n(r)(r - r_{\text{eff}})^2 \pi r^2 = \frac{r_2 + r_1}{2(r_2 - r_1)} \ln(r_2/r_1) - 1, \qquad (6)$$

respectively, where

$$\langle G \rangle = \int_{r_1}^{r_2} dr\, n(r) \pi r^2 \qquad (7)$$

is the average area of the geometrical projection per particle. The effective radius has the dimension of length and provides a measure of the average particle size, whereas the dimensionless effective variance characterizes the width of the size distribution. Using $r_{\text{eff}}$ and $v_{\text{eff}}$ as the primary size distribution parameters appears to be justified for two reasons. First, it has been demonstrated by Hansen & Travis (1974) and Mishchenko et al. (2002) that different types of size distribution (power law, log normal, gamma, etc.) having the same values of the effective radius and effective variance possess similar scattering and absorption properties, thereby making $r_{\text{eff}}$ and $v_{\text{eff}}$ convenient universal characteristics of virtually any narrow or moderately wide size distribution. Second, it would obviously be impossible to accurately retrieve the full profile of the size distribution from radar measurements at just one wavelength. Throughout this study, the effective variance is fixed at the value 0.2, which corresponds to a size distribution that is neither very narrow nor very wide. This is done to simplify the $T$-matrix computations as well as because $r_{\text{eff}}$ is a more important size distribution characteristic than $v_{\text{eff}}$. The implications of this choice of $v_{\text{eff}}$ will be discussed in Section 4. It is easy to show that for $v_{\text{eff}} = 0.2$ the constant ratio of the largest to the smallest radii of the size distribution is given by

$r_2/r_1 \approx 4.86$.

The model refractive index of the ring particles corresponds to almost pure water ice (Dones 1998; Cuzzi et al. 2002) and is equal to $1.787 + i3\times10^{-4}$ (Warren 1984).

After the single-scattering characteristics of polydisperse ring particles have been determined, the VRTE is solved for a finite, homogeneous, plane-parallel layer of random particulate medium by use of the numerically exact so-called fast invariant imbedding method (Sato et al. 1977; Mishchenko 1990). The limiting case of a semi-infinite medium is handled by solving numerically the Ambartsumian nonlinear integral equation (de Rooij 1985; Mishchenko 1996). Alternatively, the VRTE can be solved by using the adding/doubling method detailed in Hovenier et al. (2004).

The output of this computation is the diffuse Stokes reflection matrix, which is then used to find the requisite coherent reflection matrix for the exact backscattering direction according to equations (14.3.21)–(14.3.25) of Mishchenko et al. (2006). The resulting Stokes reflection matrix is the sum of the diffuse and coherent reflection matrices. The final step is to calculate the circular polarization ratio using equation (14.5.15) of Mishchenko et al. (2006).

The modeling approach outlined above can be called microphysical since it directly follows from the Maxwell equations (Mishchenko 2008). Indeed, the *T*-matrix method is explicitly based on a numerically exact solution of the Maxwell equations, while the VRTE and the theory of weak localization, as described in Mishchenko et al. (2006), are asymptotic solutions of the Maxwell equations corresponding to the limit of a very low volume density of the random particulate medium.

It is not, of course, inconceivable that the particle volume density in the A and B rings of Saturn deviates from zero significantly, especially inside wakes. However, it has been demonstrated by Mishchenko (1991) that the above theoretical approach can provide good quantitative agreement with results of controlled laboratory measurements of weak localization for particle volume densities as high as 10% (van Albada et al. 1998; Wolf et al. 1988). Furthermore, the exact computations for dense many-particle ensembles described by Mishchenko (2008) reproduce all multiple-scattering effects predicted by the low-density theories of radiative transfer and weak localization. All in all, our modeling procedure appears to be the most physically based among those ever used to analyze polarization radar observations of Saturn's rings.

Finally we note that the above approach has been used previously in extensive theoretical analyses of backscattering radar characteristics of semi-infinite and finite plane-parallel layers composed of spherical and randomly oriented nonspherical particles (Mishchenko 1992, 1996; Dlugach & Mishchenko 2006, 2007; Mishchenko et al. 2006).

## 3 NUMERICAL RESULTS

The results of our extensive numerical computations are summarized in Figs. 4–9. The Chebyshev particles are identified by their deformation parameter, while the spheroids are identified by the respective semi-axis ratio $a/b$. The results are shown in the order of increasing effective radius $r_{eff}$ separately for Chebyshev particles and spheroids. The results for Chebyshev particles with deformation parameters $\xi$ and $-\xi$ turned out to be hardly distinguishable. Therefore, we plot only the results for Chebyshev particles with positive deformation parameters. The results shown for each effective radius are intended to bracket the possible solution in terms of the particle shape and the optical thickness of the rings or to demonstrate that the solution cannot be found for any reasonable shape and optical thickness assumptions. Figure 8 also shows the values of the "diffuse" circular polarization ratio obtained by neglecting the effect of weak localization, including only diffuse multiple-scattering diagrams illustrated in Fig. 2b, and using equation (14.5.16) of Mishchenko et al. (2006). In addition, Fig. 9 shows the results of computations for an equiprobable shape mixture of prolate and oblate spheroids with aspect ratios ranging from 1.2 to 1.8. Such mixtures of particle shapes are likely to be a better representative of natural nonspherical particles than spheroids with a fixed semi-axis ratio (Mishchenko et al. 2002).

In addition to the numerical results shown in Figs. 4–9, we have also performed computations for other values of the imaginary part of the refractive index ranging from 0 to 0.01 and representing different degrees of contamination of water ice by various absorbing impurities. We have found, however, that those numerical data do not change the results of the analysis summarized below and therefore are not shown here explicitly.

Figures 4–9 illustrate well the general traits of the circular polarization ratio identified and discussed in detail in Mishchenko (1992, 1996), Dlugach & Mishchenko (2006, 2007), and Mishchenko et al. (2006). In the case of grazing incidence ($B = 0°$), the only contribution to the backscattered light comes from the first order of scattering. This means that $\mu_C$ is reduced to the so-called circular depolarization ratio $\delta_C$ of the individual ring particles, which is identically equal to zero for spherical scatterers but can significantly deviate from zero for nonspherical particles (Mishchenko and Hovenier 1995). As $B$ increases, so does the diffuse multiple-scattering contribution. In most cases the latter serves to increase the circular polarization ratio (see Fig. 8). Comparison of the diagrams in Fig. 8 with the corresponding diagrams in Figs. 4–7 shows that the additional multiple-scattering effect of weak localization is always to further increase $\mu_C$.

## 4 DISCUSSION AND CONCLUSIONS

Our analysis of the numerical results shown in Figures 4–9 is based on two fundamental premises. First, we assume that the average circular polarization ratios measured for the A and B rings of Saturn and reported in Table 2 of Ostro et al. (1980) and Table 4 of Nicholson et al. (2005) are accurate, mutually consistent, and show a real quasi-linear angular trend. Second, we assume that the ring system is not strongly stratified in the vertical direction and use the model of a vertically homogeneous plane-parallel layer of discrete random medium. This means that our conclusions may need to be revised should any of these premises turn out to be partly or completely inadequate as an outcome of future studies (Nicholson 2008). Otherwise the results of our microphysical modeling of polarized radar reflectivity of the ring system lead to the following rather far-reaching conclusions.

1. Our computations show that it is impossible to reproduce the results of radar observations theoretically without an explicit inclusion of the weak localization effect. This conclusion is well illustrated by Figure 8 and is corroborated by numerous additional computations not shown here. Of course, the effect of weak localization should have been expected to be significant since the phase angle in monostatic radar observations is by definition equal to zero (Cuzzi et al. 2002). It was important, however, to establish unequivocally that the standard model of radar reflectance based on the VRTE is completely inadequate and, thus, cannot yield a seemingly acceptable fit to the measurements for a wrong model of the rings.

2. We could not obtain a satisfactory theoretical fit by using the model of spherical ring particles. This is not surprising as the particles forming Saturn's rings consist, in all likelihood, of solid water ice and cannot be expected to be spherical. Again, however, it was important to rule out an unphysical fit based solely on the polarization radar data.

3. Our results favor the model of ring bodies in the form of nearly spherical particles with small-scale surface roughness (Chebyshev particles with $|\xi| < 0.15$) and completely rule out nonspherical particles with aspect ratios significantly exceeding 1.2 (see Fig. 4–7). All in all, our computations show that large circular polarization ratios observed for Saturn's rings are mostly the result of multiple interparticle scattering (including weak localization) rather than the result of particle nonsphericity (Mishchenko and Hovenier 1995) and/or what Nicholson et al. (2005) call multiple scattering within the large ring bodies. Admittedly, this conclusion relies, to a certain extent, on the assumed vertical homogeneity of the rings and may need to be revisited should it be firmly established that the ring system is optically thick and strongly inhomogeneous in the vertical direction.

4. Our simulations favor average ring optical thickness values in the range 2–3 or even larger. This result appears to be in reasonable agreement with recent estimates based on Cassini radio occultation observations (Marouf et al. 2007). It can be seen, however, that the retrieval of the optical thickness depends rather strongly on the assumed size of the error bars in the radar measurements and is not very well constrained.

It has been suggested recently (e.g., Daisaka et al. 2001; Porco et al. 2005; Thomson et al. 2007; Altobelli et al. 2008; Leyrat et al. 2008) that Saturn's rings are not horizontally homogeneous but rather have a pronounced wake structure with particle density between the wakes dropping to almost zero. If so, our results remain valid provided that the horizontal optical thickness of the individual wakes is significantly greater than their vertical optical thickness. In that case the retrieved optical thickness values should be attributed to an average wake rather than to the optical thickness averaged over the entire horizontal extent of the rings.

Recent studies suggest indeed that the individual wakes are geometrically thin (Hedman et al. 2007) and can be modeled as broad, flat sheets of particles with relatively empty spaces between them (Colwell et al. 2006, 2007). If so, the plane-parallel particulate layer model used in this paper appears to be quite adequate. The relatively large vertical optical thickness values favored by our model computations also appear to be in agreement with the conclusion of Colwell et al. (2007) that individual wakes can be considered nearly opaque.

It is important to realize that a wake structure can result in significant differences between the vertical optical thicknesses retrieved from analyses of radar measurements of the backscattered circular polarization ratio and those retrieved from radar measurements of backscattered intensity or radiometric measurements of directly transmitted radiation. Indeed, the former represent the individual wakes (the same filling factor $F$ representing the fraction of the total observed area occupied by the wakes enters both the numerator and the denominator of the formula defining the circular polarization ratio and thereby cancels out), whereas the latter represent the result of averaging over both the wakes and the empty areas between the wakes. This may, at least in part, explain why our vertical optical thickness values may exceed those based on the inversion of transmission observations. This also provides additional justification for our choice to analyze the (presumably more reliable) average radar data for both rings rather than those for the A and B rings separately.

Until and unless the filling factor $F$ is accurately known, the use of our model to analyze radar measurements of backscattered intensity remains highly problematic. For example, both an increase in $F$ and an increase in the average vertical optical thickness of the wakes can be expected to have the same increasing effect on the backscattered intensity. Therefore, a significant uncertainty in $F$ will invariably translate into a significant uncertainty in the retrieved wake optical thickness.

5. Our results favor particles with effective radii in the range 4–10 cm and definitely rule out effective radii significantly smaller than 4 cm. They also seem to rule out effective radii significantly greater than 10 cm. Of course, the former result does not imply that ring particles with radii much smaller than 4 cm do not exist. Indeed, it is well known that a prominent constituent of the A and B rings are sub-micrometer ice grains which cause the spectacular photometric and polarimetric opposition effects when they cover the larger ring bodies (Lyot 1929; Franklin & Cook 1965; Mishchenko & Dlugach 1992; Mishchenko 1993; Dollfus 1996; Rosenbush et al. 1997) and cause the spokes when they are levitated and create a dusty atmosphere around the larger ring particles (Doyle & Grün 1990; McGhee et al. 2005). Also, the presence of regolithic grains with a typical radius of 5–20 μm may follow from spectroscopic observations (Nicholson et al. 2008). Radar observations at 12.6 cm are completely insensitive to the presence of such small particles since the latter effectively serve as Rayleigh scatterers with a negligibly small optical thickness. This explains our modeling choice of a moderately wide size distribution with a fixed effective variance and effective radii comparable to the radar wavelength.

On the other hand, the extrapolation of the observed $\mu_C$ values to $B = 0°$ and our extensive theoretical results imply that the scattering dominance of particles with radii significantly exceeding 10 cm at the 12.6 cm radar wavelength is highly implausible. Indeed, the dashed line in Fig. 1 suggests that in the single-scattering regime ($B = 0°$) the individual ring particles must have very small depolarization ratios $\delta_C$. The only way to achieve that for ice particles with sizes significantly greater than the radar wavelength is to make their overall shape spherical and limit the amplitude of microscopic surface roughness to a small fraction of the wavelength (see the evolution of the $\mu_C(B = 0°)$ value with increasing $r_{eff}$ and/or $\xi$ in Figs. 4–6). However, such nearly spherical particles develop a strong backscattering peak in their single-scattering phase function (see the lower diagram in Fig. 9.22 of Mishchenko et al. 2002). Owing to this peak, the radar signal is dominated by the first-order scattering, which leads to the suppression of the depolarizing effect of multiple scattering. As a consequence, the theoretical $\mu_C$ values at larger opening angles remain relatively constant and cannot reach the observed values even if the effect of weak localization is fully accounted for (see the two upper right-hand diagrams in Fig. 8 and the two bottom diagrams in Fig. 6).

If one accepts the reported measurements of $\mu_C$ as real then the only conceivable way to reproduce these values with significantly larger particles might be to assume that the ring system is optically thick and strongly vertically stratified, the largest ring bodies being restricted to the mid-plane and being completely obscured by centimeter-sized particles (cf. Nicholson et al. 2005). Whether such a model of the ring system is physically plausible (cf. Cuzzi et al. 1984; Esposito 2002) and can reproduce the observed circular polarization ratios remains to be seen. Answering this question requires, in particular, complicated electromagnetic-scattering computations which are beyond the scope of this study.

The range of effective radii dictated by the radar polarization measurements may be inconsistent with the size distributions retrieved from analyses of stellar occultation observations (e.g., French and Nicholson 2000). It should be remarked, however, that the physical interpretation of such observations as well as radio occultation measurements (e.g., Marouf et al. 1982; Zebker et al. 1983, 1985) is very complicated and model-dependent (Cuzzi et al. 1984) and should include, among other optical phenomena, the effect of forward-scattering interference (Mishchenko 2008). Furthermore, unlike the backscattering polarimetric measurements, the measurements of the directly transmitted radiation represent an extremely complex convolution of the effects of particle size, interparticle distance, and the potential wake structure. The accurate and definitive de-convolution of these effects is not a straightforward task. We, therefore, believe that more studies are required in order to analyze the mutual consistency of all available datasets (including the existing radar measurements of the circular polarization ratio) and to develop a physically based model of Saturn's rings capable of reproducing simultaneously all observations that are deemed reliable.

**ACKNOWLEDGMENTS**

We thank P. D. Nicholson for very useful comments which resulted in a much improved manuscript. This research was sponsored by the NASA Radiation Sciences Program managed by Hal Maring.

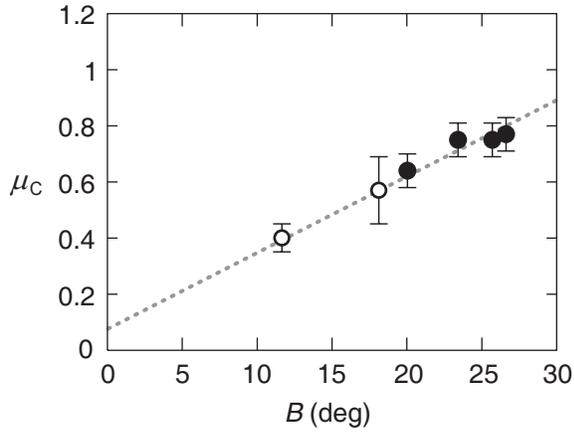

**Figure 1.** Variation of the average polarization ratio $\mu_C$ with ring opening angle $B$ for the A and B rings of Saturn based on all available radar data at 12.6 cm. Open circles show data from Table 2 of Ostro et al. (1980), while filled circles show the results from Table 4 of Nicholson et al. (2005).

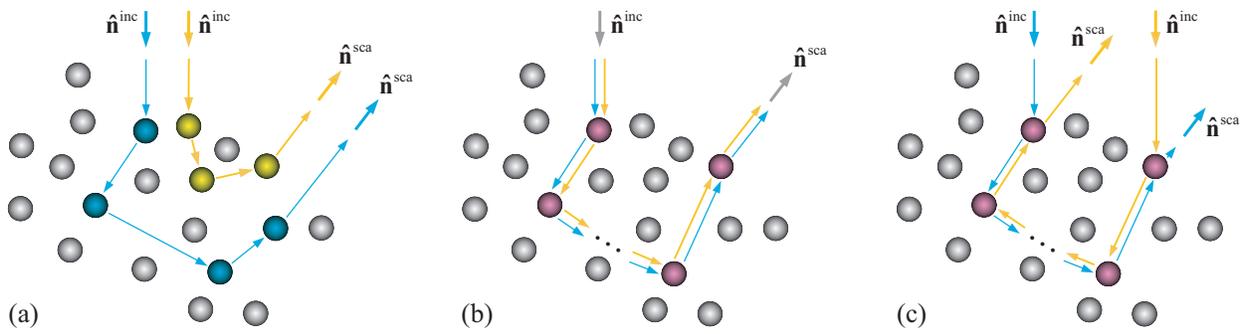

**Figure 2.** (a) Interference origin of speckle. (b) Interference origin of the diffuse background. (c) Interference origin of weak localization.

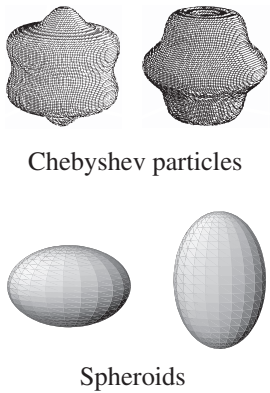

Chebyshev particles

Spheroids

**Figure 3.** Upper panel: Chebyshev particles with $\xi = 0.1$ (left) and $-0.1$ (right). Lower panel: oblate spheroid with $a/b = 3/2$ (left) and prolate spheroid with $a/b = 2/3$.

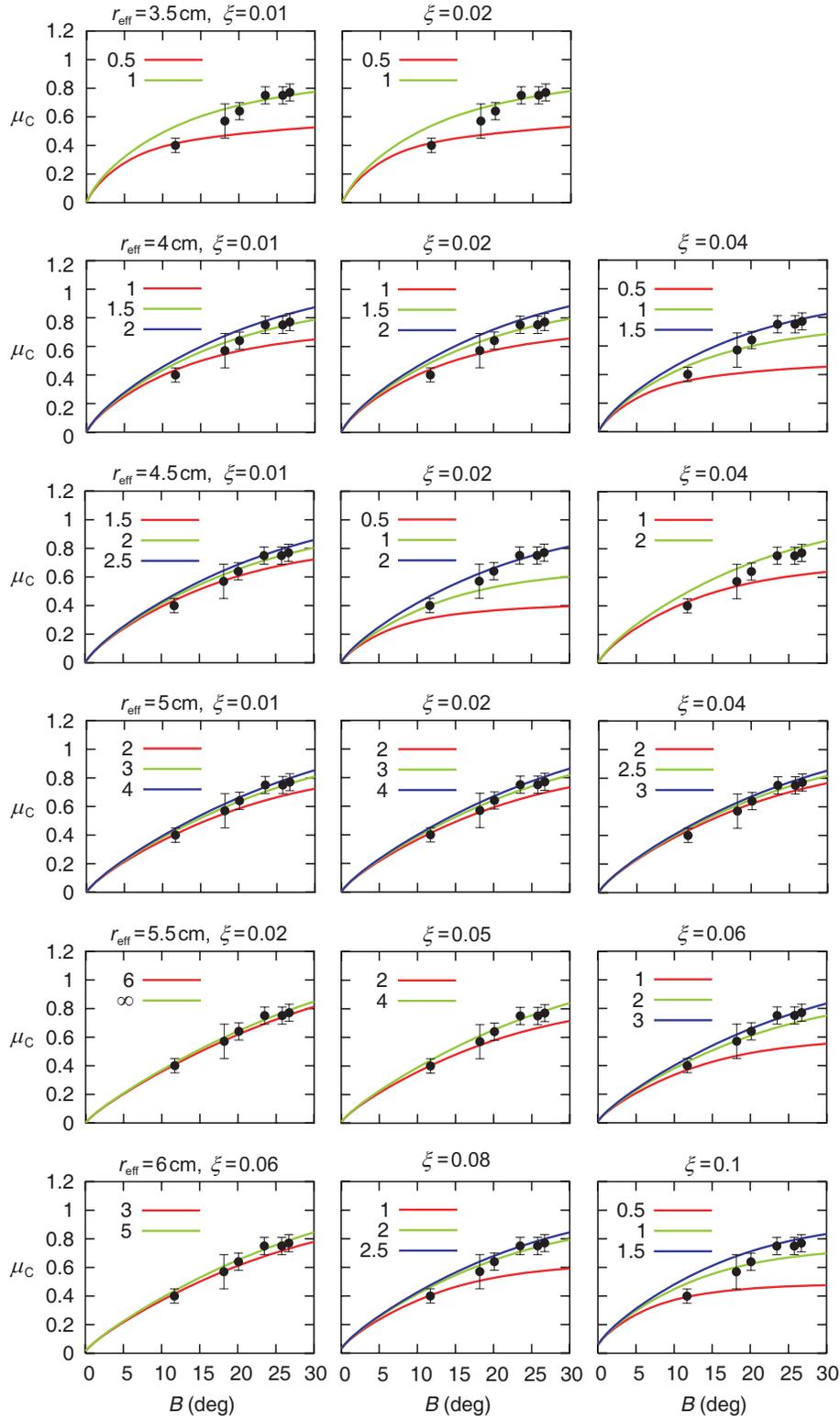

**Figure 4.** Circular polarization ratio versus ring opening angle for a plane-parallel layer consisting of polydisperse, randomly oriented Chebyshev particles with effective radii ranging from 3.5 to 6 cm. Different colors correspond to different optical thicknesses of the layer, as indicated in the insets.

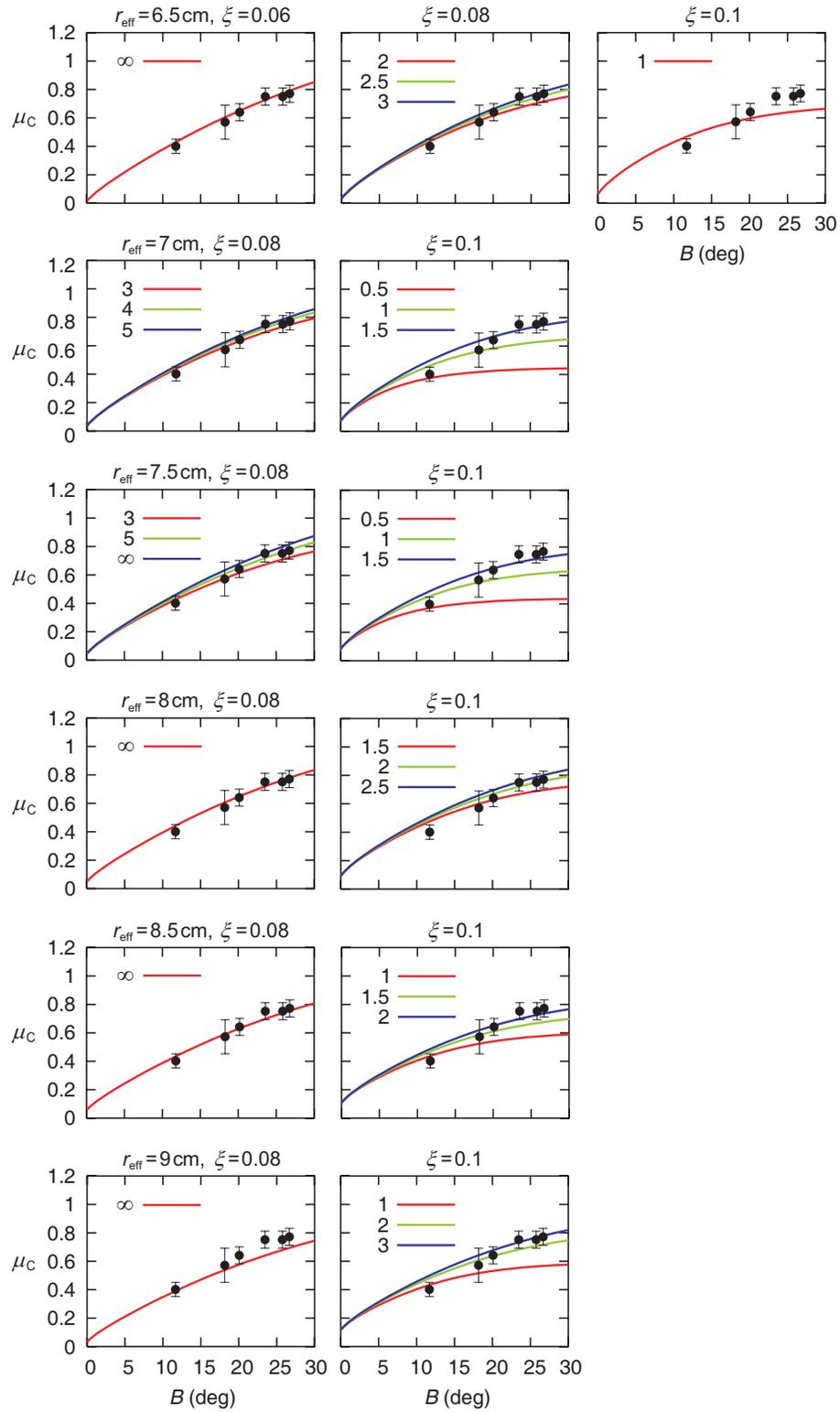

**Figure 5.** As in Fig. 4, but for effective radii ranging from 6.5 to 9 cm.

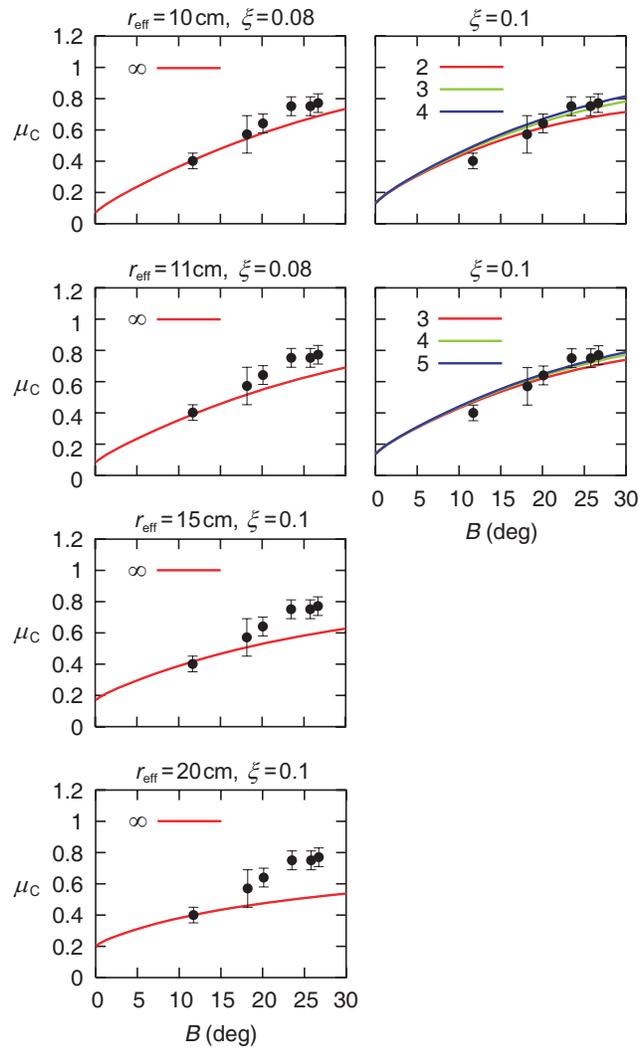

**Figure 6.** As in Fig. 4, but for effective radii ranging from 10 to 20 cm.

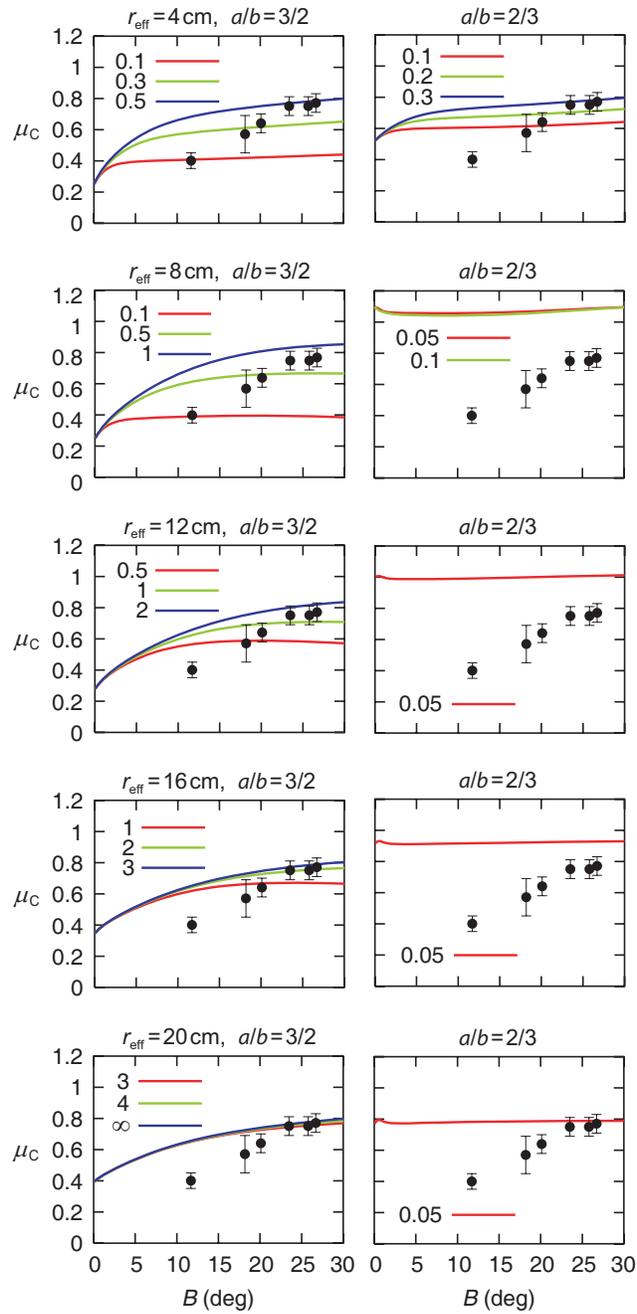

**Figure 7.** As in Figure 4, but for polydisperse, randomly oriented spheroids with effective radii ranging from 4 to 20 cm.

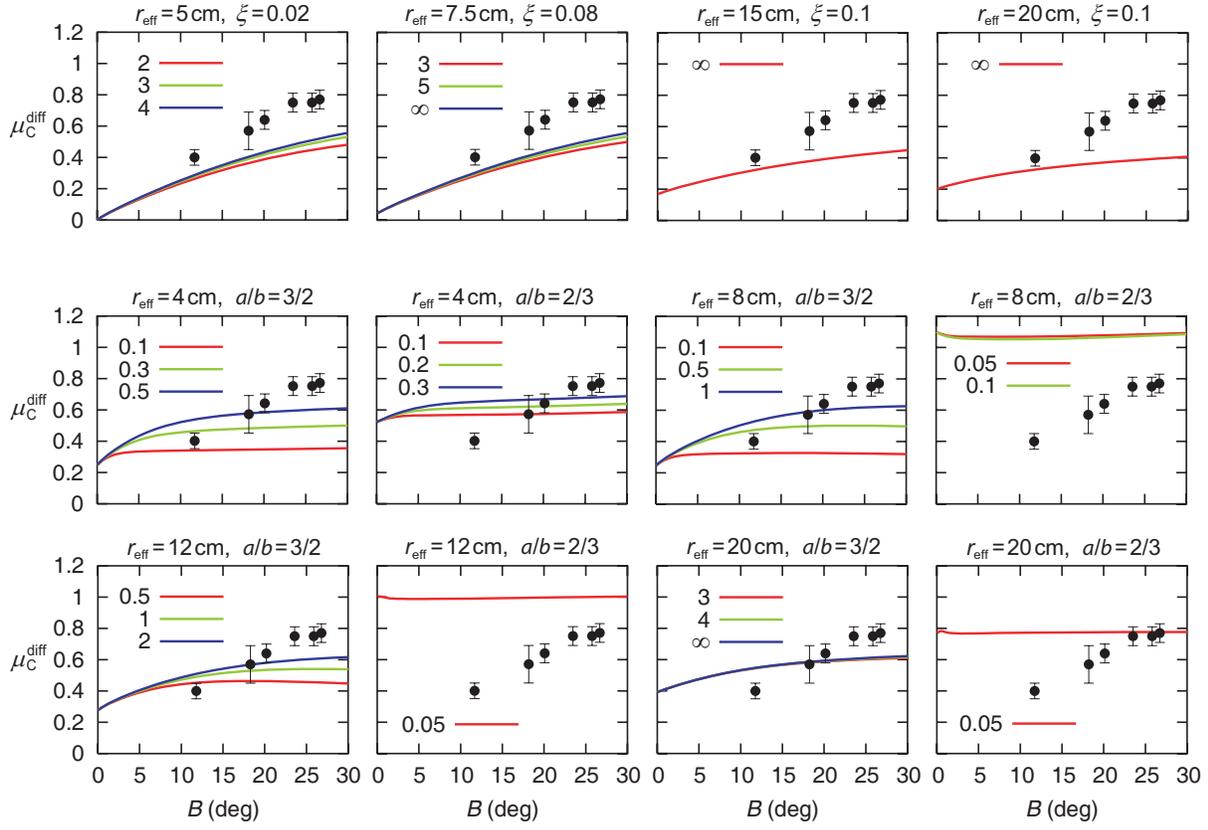

**Figure 8.** Diffuse circular polarization ratios computed for polydisperse, randomly oriented Chebyshev particles (top row) and prolate and oblate spheroids (two bottom rows).

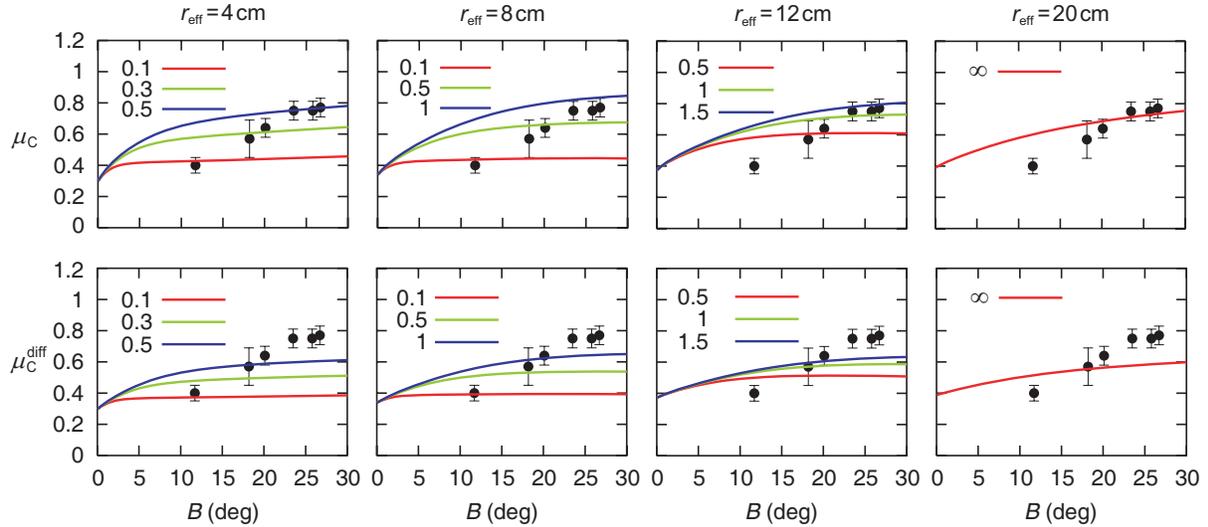

**Figure 9.** Circular (top row) and diffuse circular (bottom row) polarization ratios computed for an equiprobable shape mixture of polydisperse, randomly oriented prolate and oblate spheroids with aspect ratios ranging from 1.2 to 1.8.

18